\documentclass[aps,showpacs,twocolumn]{revtex4}
\usepackage{amsmath}
\input{epsf.tex}
\epsfverbosetrue

\begin{document}

\input epsf
\newcommand{\infig}[2]{\centerline{\epsfxsize=#2\columnwidth \epsfbox{#1}}}
\newcommand{\be}{\begin{equation}}
\newcommand{\nn}{\nonumber}
\newcommand{\ee}{\end{equation}}
\newcommand{\bea}{\begin{eqnarray}}
\newcommand{\eea}{\end{eqnarray}}
\newcommand{\bin}[2]{\left(\begin{array}{c} #1 \\ #2\end{array}\right)}
\newcommand{\pred}{^{\mbox{\small{p}}}}
\newcommand{\retr}{^{\mbox{\small{r}}}}
\newcommand{\retd}{_{\mbox{\small{r}}}}
\newcommand{\p}{_{\mbox{\small{p}}}}
\newcommand{\m}{_{\mbox{\small{m}}}}
\newcommand{\tr}{\mbox{Tr}}
\newcommand{\rs}[1]{_{\mbox{\small{#1}}}}	
\newcommand{\ru}[1]{^{\mbox{\small{#1}}}}
\newcommand{\ris}[2]{_{\mbox{\small{#1}}{#2}}}	
\title{Fidelity for imperfect postselection}
\author{Craig S. Hamilton and John Jeffers}
\affiliation{Computational Nonlinear and Quantum Optics Group, SUPA, Department 
of Physics, University of Strathclyde, John Anderson Building, 107 Rottenrow, 
Glasgow G4 0NG, U.K.}
\pacs{42.50.-p 42.79.-e 03.65.Ta}

\begin{abstract}
We describe a simple measure of fidelity for mixed state postselecting devices. 
The measure is most appropriate for postselection where the task performed by 
the output is only effected by a specific state. 
\end{abstract}
\maketitle
\section{Introduction}
\begin{figure}[phtb]
\label{postd}
\centerline {\epsfxsize=8.0cm \epsffile{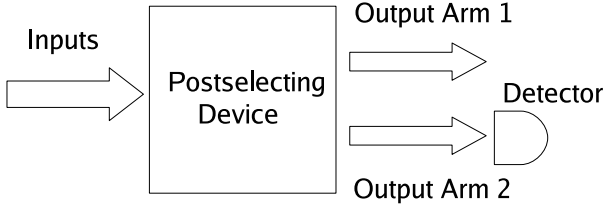}}
\caption {Typical postselecting device}\label{bs}
\end{figure}

Fidelity provides a measure of the closeness of two states. One situation to 
which it applies directly is postselection, in which a postselecting device 
(Fig.~\ref{bs}) produces a quantum state conditioned on the result of a 
measurement. Postselection is a useful technique for producing particular 
quantum states, for example in linear optical quantum computing (LOQC) 
\cite{klm,loqc} and in other cluster state or matter-based schemes 
\cite{Nielsen, Browne, Joo, Louis, Kok}. If either the measuring device or the internal components of the postselector are imperfect then the state produced 
may not be that which was intended. 
In most cases postselectors are designed to produce pure states when 
functioning correctly, but this is not always the case, and we shall see that 
the concept of mixed state fidelity can be applied to postselection with 
imperfect internal components. 

A standard measure of fidelity comparing mixed states $\hat{\rho}_a$ and 
$\hat{\rho}_b$ is $F=\{\tr [ (\hat{\rho}_a^{1/2} \hat{\rho}_b 
\hat{\rho}_a^{1/2})^{1/2} ]\}^2$, which reduces to $F=\tr\left( \hat{\rho}_a 
\hat{\rho}_b \right)$ if either of the two states (say $\hat{\rho}_b$) is pure \cite{Jozsa}. In this case the 
measure is effectively the probability that the state $\hat{\rho}_a$ would pass 
a measurement test with $\hat{\rho}_b$ as one of the outcomes. Others use the 
square root of this quantity as a measure of fidelity \cite{Fuchs,Nielsen00}. 
Consider a postselector designed to produce a pure state when functioning 
perfectly, with perfect internal components but an imperfect detector. For such 
a device it has previously been shown that the retrodictive conditional 
probability that the detector correctly indicates the detector arm state is a 
close lower bound to the pure state fidelity above \cite{Jeffers06}. This 
measure, the retrodictive fidelity, is the most appropriate measure of fidelity 
when the only useful output state of the device is the one which it would 
produce if it were working perfectly. Its advantages are that (i) it is simple 
to calculate because it depends only on detector arm properties, and not 
explicitly on the actual state produced by the device, and (ii) it is the 
natural quantity to attempt to maximise to improve the fidelity in an 
experiment. Recently the measure has been used to show that placing an optical 
amplifier in front of an imperfect photodetector can greatly improve the 
fidelity based on the detector results \cite{Jeff07}. The method works best for 
postselection based on zero photocounts at the detector. 

In this paper we generalise the results found in \cite{Jeffers06} to mixed 
state fidelities, and apply them to three practical situations. The paper is 
organised as follows. In section 2 we derive a mixed state fidelity, which we 
have called the correct output fidelity $F\rs{c}$, that is most appropriate for 
postselection when a particular state is the only one which will perform the 
quantum information task required of the postselected output.  In section 3 we 
apply this measure to the practical situation of the comparison of two coherent 
states selected at random from a limited set. We show that we can increase the 
confidence in the comparison by increasing $F\rs{c}$ with preamplified detectors. 
Next we apply the measure to a lossy beam splitter, using two practical examples. Firstly we look at two photon state generation, and secondly we apply the measure to the nonlinear sign-shift gate, which uses two such beam splitters.
Finally we summarise our results and conclude. 

\section{Mixed state fidelity for imperfect postselectors}
In this section we derive the correct output fidelity for a postselecting 
device which is required to produce a particular state. The calculation is a 
generalisation to mixed states of the results which appear in \cite{Jeffers06} 
for pure state postselectors. The postselector is here assumed to be imperfect 
even if the detection on which it is based is perfect. 

A typical postselector (Fig.~\ref{bs}) has at least two output arms, the joint 
output state of which will be entangled. Therefore measuring the state in one 
arm can collapse the state in the other arm. Typically in optics the 
measurement device will be a photodetector, and the measurement will be in the 
photon number state basis. This situation allows the engineering of almost any 
finite superposition of number states at the output even if the inputs to the 
device are infinite superpositions of number states \cite{Pregnell}. 
Unfortunately the optical components which make up the postselector, and the 
detectors, will be imperfect, and so the device will not produce the 
postselected states advertised by the detection results. 
\subsection{Fidelity for perfect detection} 
The joint output state of the two arms of the postselector is 
$\hat{\rho}_{12}$. Suppose that the (perfect at this stage) detector in arm 2 
has a set of orthogonal measurement results  $\hat{\pi}_2^m$ (corresponding, 
for example, to numbers of photocounts in a photodetector). If the result $m$ 
is obtained this corresponds to measuring arm 2 to have been in the state 
$\hat{\rho}_2^m=\hat{\pi}_2^m/\mbox{Tr}\hat{\pi}_2^m$. This measured state 
provides a valid description of the arm 2 state in retrodictive quantum theory, 
in which the measured state evolves backwards in time \cite{retqth}. When the 
result $m$ is obtained the device produces the arm 1 state
\bea
\label{rho1m}
\hat{\rho}_1^m= \frac{\tr_2 \left(\hat{\rho}_{12}\hat{\pi}_2^m\right)}
{\tr_{12} \left(\hat{\rho}_{12}\hat{\pi}_2^m\right)}.
\eea
As the detector is insensitive to any off-diagonal elements in the detection 
basis we can rewrite the mode 2 state $\hat{\rho}_2=\tr_1 \hat{\rho}_{12}$ as 
\bea
\label{lambda2}
\hat{\Lambda}_2 = \sum_m p_m \hat{\rho}_2^m,
\eea
where $p_m$ is the probability of obtaining detection result $m$ for a perfect detector, or loosely, the prior probability that the arm 2 state is 
$\hat{\rho}_2^m$. Then Eq. (\ref{rho1m}) becomes
\bea
\hat{\rho}_1^m= \frac{\tr_2 \left(\hat{\rho}_{12}\hat{\pi}_2^m\right)}
{\tr_2 \left(\hat{\Lambda}_2\hat{\pi}_2^m\right)}.
\eea
When the particular result $n$ is obtained at the detector we assume that the postselector has worked, and so the state $\hat{\rho}_1^n$ is produced by the 
device. The postselector is assumed to be imperfect even when the detection is 
perfect; it does not produce the exact required state. Suppose that the arm 1 
state that is required is the correct state $\hat{\rho}_c$. If both 
$\hat{\rho}_1^n$ and $\hat{\rho}_{c}$ are mixed then we can use the general 
form of the fidelity \cite{Jozsa}
\begin{eqnarray}
\label{F}
F=\left\{\tr \left[ \left(\hat{\rho}_c^{1/2} 
\hat{\rho}_1^n \hat{\rho}_c^{1/2}\right)^{1/2} \right]\right\}^2.
\end{eqnarray}
We can write $\hat{\rho}_1^n$ as follows
\bea
\label{rho1decomp}
\hat{\rho}_1^n = P\ru{max} \hat{\rho}_c + \hat{\gamma},
\eea
where $P\ru{max}$ is the maximum fraction of $\hat{\rho}_c$ that can be `made' 
from $\hat{\rho}_1^n$, and $\hat{\gamma}$ is the remainder, with positive 
diagonal elements in any basis
\footnote{As a simple example consider the two states 
$\hat{\rho}_a = \frac{1}{3}\left( |1\rangle \langle1| +|2\rangle \langle2| 
+|3\rangle \langle3|  \right)$ and  $\hat{\rho}_a = \frac{1}{3} 
\left( 2|2\rangle \langle2| +|3\rangle \langle3|  \right)$. We can write 
$\hat{\rho}_a = \frac{1}{2} \hat{\rho}_b +\frac{1}{3} |1\rangle \langle1|
+ \frac{1}{6}|3\rangle \langle3|$, and so the fraction $P\ru{max} = 1/2$ in 
this case. We cannot write $\hat{\rho}_b$ in terms of $\hat{\rho}_a$ for any 
positive remainder.}. In optics such a decomposition will normally be possible for 
pairs of states which are finite (pure or mixed) sums of number states. For 
continuous variable schemes, where the basis states are coherent or squeezed 
such decompositions may not always be possible due to the fact that the number 
state decompositions contain an infinity of terms. We can derive a lower bound 
on the fidelity $F$, the correct output fidelity $F\rs{c}$ by substituting only the 
first term in Eq. (\ref{rho1decomp}) into Eq. (\ref{F})
\bea
F\rs{c}=P\ru{max}.
\eea
Although this is a lower bound on $F$, when only the correct state 
$\hat{\rho}_c$ performs the appropriate quantum information task at the output, 
$F\rs{c}$ and not $F$ is the appropriate measure of fidelity. Note that $F\rs{c}$ is 
not the probability of passing a measurement test, but is in a loose sense the probability that the output state ``is" $\hat{\rho}_c$.

\subsection{Imperfect detection}
Here we follow the method introduced in \cite{Jeffers06} to include the effects 
of imperfect detection. In this case obtaining the measurement result $n$ does 
not correspond perfectly to the probability operator $\hat{\pi}_2^n$. Instead 
it corresponds to a mixture of all of the possible $\hat{\pi}_2^m$ 
\cite{mixed}, given by
\bea
\hat{\pi}_2^{\prime n} = P\pred(n|n) \hat{\pi}_2^n + \sum_{m\neq n} P\pred(n|m) 
\hat{\pi}_2^m
\eea
where $P\pred(m|n)$ is the predictive conditional probability that state 
$\hat{\rho}_2^n$ gives result $m$ at the detector \cite{Jeffers06}. The state 
produced by the postselector is then found from Eqs. (\ref{rho1m}) and 
(\ref{lambda2}) to be 
\bea
\label{rhoprimed}
\hat{\rho}_1^{\prime n}= \frac{\tr_2 \left(\hat{\rho}_{12}
\hat{\pi}_2^{\prime n}\right)}
{\tr_{2} \left(\hat{\Lambda}_2\hat{\pi}_2^{\prime n}\right)} 
= P\retr(n|n)\hat{\rho}_1^n +\sum_{m\neq n}P\retr(m|n)\hat{\rho}_1^m,
\eea
where Bayes' theorem has been used to write the result in terms of the 
retrodictive conditional probabilities that particular states were present in 
the measurement arm given the detection result $n$. 

The density operators which appear in the second term of Eq. (\ref{rhoprimed}) 
are those which would have been produced by the device if the detector had been 
perfect, and a result other than $n$ had been obtained. Usually these will have 
small overlap with the required state. In any case $P\retr(n|n)\approx 1$ for a 
sufficiently good detector and $P\retr(m\neq n|n)\approx 0$. Thus the second 
term in Eq. (\ref{rhoprimed}) is small, and we neglect it from here on. In 
calculating the fidelity, instead of $\hat{\rho}_1^n$ we use the first term in 
$\hat{\rho}_1^{\prime n}$, and find that the correct output fidelity is
\bea
F\rs{c}=P\ru{max}P\retr(n|n).
\eea
$F\rs{c}$ depends on two factors. The first is output state specific, and the 
second depends explicitly only on detector arm properties. The second factor 
has been dubbed the retrodictive fidelity \cite{Jeffers06}. It has been 
recently used as a fidelity quantifying the conditional preparation of number 
states \cite{Rohde}. Each can be calculated independently from simple 
properties of both the device for perfect detection and the imperfect detector. 
This makes the correct output fidelity much easier to calculate than Eq. 
(\ref{F}) in many cases. It also makes clear how improving the postselector 
increases the fidelity. Either the postselecting system can be improved, by for 
example using better components, or the confidence in the detection can be 
improved \cite{Croke06}. 

\section{Comparison of Coherent States}
Recently a quantum key distribution protocol has been proposed 
\cite{Andersson06} for sending a key to more than one recipient. The protocol 
relies on the comparison of coherent states chosen at random from a finite set. 
If the coherent states passed to each recipient are equal, this means that each 
recipient receives the same key bit. The comparison offers some protection 
against either the key sender or the recipients deliberately corrupting the 
key, and against eavesdropping. In order to determine if two coherent states 
are identical a test must be passed. The simple test devised in 
\cite{Andersson06} is that if equal coherent states form the inputs to a 50/50 
asymmetric beam splitter from separate arms then one of the outputs will be in 
a vacuum state: no photocounts can be recorded there. However, when different 
coherent states are input there will be a non-zero coherent state in the 
detector arm, and photocounts can be recorded. The advantage of using coherent 
states is that they can be compared almost non-invasively. If the measurement 
arm state is the vacuum state we can re-obtain the initial states from the 
other arm simply by passing the unmeasured output through a further 50/50 beam 
splitter. 

\begin{figure}[phtb]
\centerline {\epsfxsize=8.0cm \epsffile{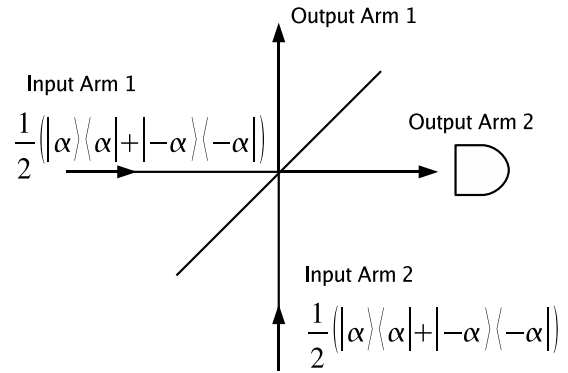}}
\caption{Beam splitter with coherent state inputs. There is a phase change of 
$\pi$ on reflection from arm 2.}\label{cbs}
\end{figure}

In order to compare the two coherent states to determine if they are identical 
they are sent through a beam splitter as in Fig\.~\ref{cbs}. The transformation 
for a coherent state passing through a beam splitter is straightforward
\cite{Loudon}, and for two arbitrary coherent states, $|\alpha \rangle$ and $|\beta \rangle$, falling on an asymmetric 
beam splitter with a reflection phase change of $\pi$ from one arm it is 
\begin{equation}\label{coh_bs_transform}
|\alpha\rangle \langle \alpha | \otimes |\beta \rangle \langle \beta| \rightarrow
|t\alpha+r\beta\rangle\langle t\alpha+r\beta|\otimes |t\beta-r\alpha\rangle\langle
t\beta-r\alpha|
\end{equation}

Suppose that in each input arm a choice of one of the two coherent states 
$| \alpha \rangle$ and $ | -\alpha \rangle$ is made randomly. Then the 
{\it a priori} density operator in each arm is the mixture indicated in 
Fig. (\ref{cbs}). The beam splitter transmission and reflection coefficients 
are $1/ \sqrt 2$, with a phase change of $\pi$ on reflection from arm 2. The 
output state is then

\bea
\nn \hat{\rho}_{12}&=&\frac{1}{4}\left[ \left(| \sqrt{2} \alpha \rangle_{1}\langle
\sqrt{2} \alpha| + | -\sqrt{2} \alpha \rangle_{1}\langle -\sqrt{2} \alpha| 
\right)\otimes | 0 \rangle_{2}\langle 0| \right. \\
\nn &+& \left. |0  \rangle_{1} \langle 0| \otimes  \left ( |-\sqrt{2}\alpha \rangle_{2}\langle -\sqrt{2} \alpha | + |\sqrt{2}\alpha \rangle_{2}\langle \sqrt{2} \alpha |\right) \right].\\
\rule{0mm}{0mm}
\eea

If we place a photodetector in arm 2 we can view the device as a mixed-state 
postselector, which postselects on the basis of recording no counts at the detector. If the input states are identical then the vacuum state is 
produced in arm 2 and the arm 1 state is the mixed state 
\bea
\label{rho1c}
\hat{\rho}_1=\frac{1}{2} \left(| \sqrt{2} \alpha \rangle_{1}\langle
\sqrt{2} \alpha| + | -\sqrt{2} \alpha \rangle_{1}\langle -\sqrt{2} \alpha| 
\right).
\eea 
This state forms the correct output state $\hat{\rho}_c$.
However, if the two input states are different then the arm 2 output state is 
\bea
\hat{\rho}_2=\frac{1}{2} \left(| \sqrt{2} \alpha \rangle_{2}\langle
\sqrt{2} \alpha| + | -\sqrt{2} \alpha \rangle_{2}\langle -\sqrt{2} \alpha| 
\right),
\eea 
and the arm 1 output is the vacuum. The nonorthogonality of the coherent and 
vacuum states means that if no counts are recorded in arm 2 it is possible that 
the arm 2 state is the coherent state mixture above. Thus the postselector is 
not perfect, even if the detector is. 

When no counts are recorded in arm 2 the measurement operator is 
$\hat{\pi}_2^0=|0 \rangle_2 \langle 0 |$ and we now have the state in output 
arm 1,  
\begin{eqnarray}\label{rho1}
\nn \hat{\rho}_1^{n=0} &=&  \frac{1}{2\left(1+e^{-2\left|\alpha \right|^2}\right)}\left[| \sqrt{2} \alpha \rangle \langle 
\sqrt{2} \alpha| \right.\\
&+& \left. | -\sqrt{2} \alpha \rangle \langle -\sqrt{2} \alpha| + 
2e^{-2\left| \alpha \right|^2}| 0 \rangle \langle 0| \right]
.
\end{eqnarray}
We can use the fidelity of this produced state in relation to the correct state given by Eq. (\ref{rho1c}) as a measure of the quality of the comparison. We first consider the correct output fidelity $F\rs{c}$ leaving comparison with $F$ in this system to later.  
We can can write (\ref{rho1}) as,
\begin{eqnarray}
\hat{\rho}_1^{n=0} = P\ru{max}\hat{\rho}_c +\hat{\gamma}
\end{eqnarray}
where $P\ru{max}=1/(1+e^{-2|\alpha|^2})$,
$\hat{\rho}_c=\frac{1}{2}[| \sqrt{2} \alpha \rangle \langle \sqrt{2} \alpha|+
\\|-\sqrt{2} \alpha \rangle \langle -\sqrt{2} \alpha|]$
and $\hat{\gamma}=|0\rangle \langle 0| e^{-2\left|\alpha \right|^2}/(
1+e^{-2\left|\alpha \right|^2})$. The correct output fidelity is then
\begin{eqnarray}
\label{fc}
F\rs{c}=P\ru{max}=\frac{1}{1+e^{-2\left|\alpha\right|^2}}.
\end{eqnarray}
This quantity sets a limit on the distinguishability of the two input coherent 
states given no counts. It tends to unity for large values of $|\alpha|$, in 
line with our expectation, as the coherent states become more orthogonal. 

\subsection{Imperfect detection and improving the comparison}

Imperfect photodetection will degrade the fidelity further. A photodetector 
with a poor quantum efficiency will have an increased probability of a readout 
of zero photocounts, as sometimes when there is one photon `present' in the 
measurement arm it will be registered as zero counts. Then the state produced 
by the device will be the incorrect one. In the comparison of coherent states 
the effect of this is that the comparison test will be passed more often than 
it should be. Reduced detector efficiency lowers the confidence in the 
comparison. For this reason discarding a correct state (a false negative error) 
is not as damaging to the key distribution protocol as accepting an incorrect
state (a false positive error). 

\begin{figure}[phtb]
\centerline {\epsfxsize=7.0cm \epsffile{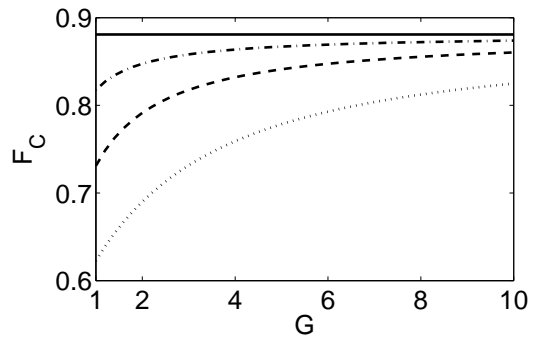}}
\caption{Fidelity vs. amplifier gain for four values of detector quantum efficiency $\eta$. From top to bottom the curves are for $\eta=1, 0.75, 0.5, 0.25$.}\label{fid-vs-amp}
\end{figure}

One method which helps to distinguish between the vacuum state and other number states 
when we have a lossy photodetector is to place an amplifier in front of the 
photodetector \cite{Jeff07}. Noiseless amplification is possible classically, but in quantum physics the process is accompanied by the addition of extra photons not associated with the amplified input \cite{Caves82}. Despite this, it has been shown that for quantum systems amplification and attenuation are inverse processes 
\cite{Ottavia04, Barnett00}, contrary to usual belief in quantum optics. In 
order to model this situation we modify our measurement operator to include the 
effects of non-unit efficiency detection, which is equivalent to placing an 
attenuator in front of a perfect detector \cite{Shapiro,Loudon}, and 
amplification. Our perfect measurement operator, $\hat{\pi}^0$ now becomes,
\begin{equation}
\hat{\pi}^{0'}=\sum_{n=0}^{\infty}\frac{(1-\eta)^n}{G^{n+1}}\sum_{m=0}^{n} 
\left (\begin{array}{c} n \\ m \end{array}\right)(G-1)^{n-m}\hat{\pi}^m
\end{equation}
where $\eta$ is the quantum efficiency of the detector, the probability that 
the detector measures an individual photon that enters, $G$ is the gain of the 
amplifier and $\hat{\pi}_n$ is the projector onto the $n\ru{th}$ number state. 
We assume that the amplifier is as good as is allowed by quantum theory, adding 
the minimum average number of noise photons. The state in output arm 1, when we 
measure no photocounts, is now given by
\begin{equation}
\label{rho1eta}
\hat{\rho}_1^{n=0'}=\frac{\mbox{Tr}_2[\hat{\rho}_{12}\hat{\pi}^{0'}]}
{\mbox{Tr}_2[\hat{\Lambda}_2\hat{\pi}^{0'}]}.
\end{equation}
From this and Eq. \ref{rhoprimed}, $P\retr(0|0)$, the retrodictive conditional probability is expressible in terms of a quotient of series which can be summed to give
\begin{equation}
\label{pr00}
P\retr(0|0)=\frac{ 1+ e^{-2|\alpha|^2} }{1+ e^{-2|\alpha|^2\{\eta G/[1-\eta(G-1)]\}}}.
\end{equation}
The correct output fidelity, $F\rs{c}$, is then given by the product of Eqs. 
(\ref{pr00}) and (\ref{fc}), 
\bea
F\rs{c} = P\ru{max} P\retr(0|0) = \frac{ 1 }{1+ e^{-2|\alpha|^2\{\eta G/[1-\eta(G-1)]\}}},
\eea
which we have plotted in Fig. (\ref{fid-vs-amp}) 
for four different values of $\eta$ and for $|\alpha|=1$. It can be seen 
immediately from the graph that an amplifier will improve the fidelity of the 
postselecting device for a lossy detector. This method gives no improvement for 
a perfect detector, as expected.

\subsection{Photocount probability and information}

\begin{figure}[phtb]
\centerline {\epsfxsize=7.0cm \epsffile{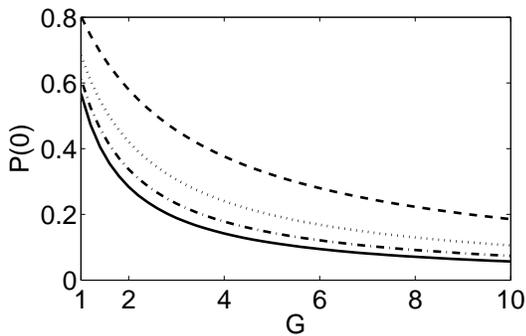}}
\caption {Probability of recording zero photocounts vs. Amplifier gain. From top to bottom the curves are for $\eta=0.25, 0.5, 0.75, 1$.}\label{P0gain}
\end{figure}

The downside to this method for improving the fidelity is that the amplifier 
reduces the probability of obtaining zero photocounts. Although when we obtain 
no counts we can be more certain that the vacuum state was present in the 
measurement arm, we measure this less often, as an amplifier adds noise photons 
to the system. This can be seen in Fig. (\ref{P0gain}), which shows the 
probability of no counts being measured by a lossy detector for the same four 
values of $\eta$.

\begin{figure}[phtb]
\centerline {\epsfxsize=7.0cm \epsffile{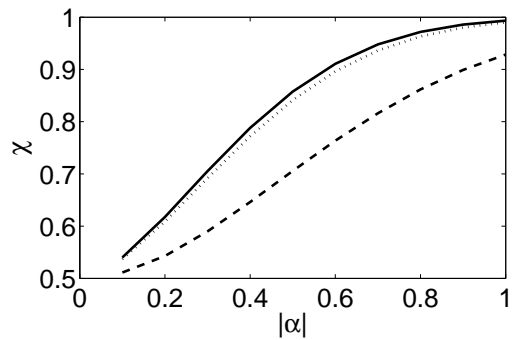}}
\caption {Accessible information in the quantum state vs. coherent amplitude. The top, full, line is 
the accessible information contained in a mixture of 3 states: $|\alpha\rangle$, $|-\alpha\rangle$ 
and the vacuum state. The middle, dotted line is the information contained in the state after 
attenuation only ($\eta=0.9$). The lower line is for the same state after being passed through an 
amplifier (G=1.5) and attenuator($\eta=0.9$).}\label{accin}
\end{figure}

One concern in transmitting quantum states is that the information carried in 
those states should not approach the classical limit, because then an 
eavesdropper could intercept the key and copy it. A transmitted coherent state 
approaches this limit when its amplitude, $|\alpha|$, becomes too large. 
Although it increases the confidence in the detection results amplification 
should not increase the information present. The fact that the detection 
probability decreases bears this out. A more quantitative check 
can be performed by noting that the accessible information contained in a 
quantum state has an upper bound given by the quantity \cite{Nielsen} 
\begin{equation}
\chi=S(\hat{\rho})-\sum_{i}p_i\hat{\rho}_i
\end{equation}
Where $\hat{\rho}=\sum_ip_i\hat{\rho}_i$ and $S(\hat{\rho})$ is the Von Neumann 
entropy of that quantum state. We compare in Fig. (\ref{accin}) the information 
held in the quantum state
\bea
\hat{\rho}=\frac{1}{2}|0\rangle \langle 0| +\frac{1}{4}(|\alpha\rangle\langle\alpha|+
|-\alpha\rangle\langle-\alpha|)
\eea
before and after it has been amplified and attenuated. We examine this state in 
particular as it is of the same form as the state transmitted along the 
detector arm in our postselecting device shown 
in Fig. (\ref{cbs}). In Fig. (\ref{accin}) the top line is the information contained in the state 
without amplification, the middle line is attenuation of the state only and the bottom line is the 
same state after passing through an amplifier and attenuator. From the graph is can be seen that 
passing the quantum state through either an amplifier, an attenuator or both decreases the 
accessible information in the quantum state. This is expected as the actions of the amplifier and 
the attenuator reduce the coherence in the quantum state and therefore decrease the possible 
information that can be carried. It can be also be seen that the accessible information tends to the 
classical limit as the coherent state magnitude tends to infinity. 

Although the decrease in accessible information is significant even for modest gain this is not such 
a problem. In a quantum communication system confidence in the result is the most important criterion. 
Provided that there is not too much excess amplifier noise the correct output fidelity, and hence the 
confidence will always be increased by amplification. The effect of detector dark counts is more 
subtle, and can either increase or decrease the confidence. 

\subsection{Comparison with standard mixed state fidelity}

\begin{figure}[phtb]
\centerline {\epsfxsize=7.0cm \epsffile{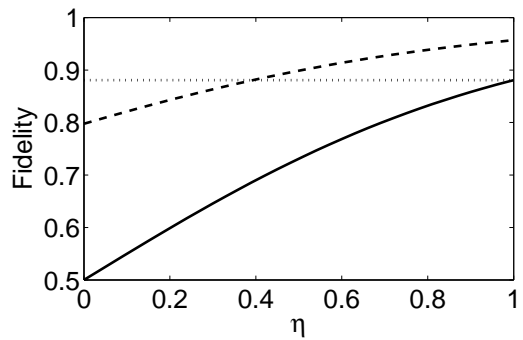}}
\caption {Correct output fidelity (solid) and standard mixed state fidelity (dashed) as a function of detector quantum efficiency. The horizontal dotted line corresponds to $P\ru{max}$}\label{fidcomp}
\end{figure}

The correct output fidelity will always be a lower bound to the standard mixed state fidelity. How much lower it is depends on the overlap of the discarded portions of the output with the desired output state. These portions will include contributions from all of the terms in Eq. (\ref{rhoprimed}), not simply the first. For this reason the standard mixed state fidelity is more complicated to calculate. Fig. (\ref{fidcomp}) compares the correct output fidelity with fidelity defined by Eqs. (\ref{F}), (\ref{rho1c}) and (\ref{rho1eta}), as a function of detector quantum efficiency, for the coherent state comparison system. Varying $\eta$ in effect varies the mixedness of the output state produced by the postselector.
The correct output fidelity is significantly lower, and is bounded by $P\ru{max}$. The fidelity of Eq. (\ref{F}) is above this value for sufficiently good detectors. The reason is the inclusion in it of terms corresponding to two effects. Firstly it includes the effect of the vacuum component of the coherent state in the detector arm, which corresponds to incorrect functioning of the device. Secondly it should be emphasised that for perfect detectors any nonzero number of counts in the detector arm corresponds to a coherent state, and not the vacuum state, being present in that arm. If the detector is lossy, sometimes it will record no counts even though it ought to have recorded some. The inclusion of terms corresponding to these two situations renders the standard mixed state fidelity insecure here. 
\section{Lossy beam splitter}
\subsection{Two-photon state generation}
A second example shows the applicability of the concept to another simple system. It is well-known that 
when two single photons interact at a 50/50 beam splitter the phenomenon known as two-photon interference 
causes them to leave by the same output port \cite{Loudon,Fearn,Mandel}. The effect is the basis of gate 
operations in linear optical quantum computing \cite{klm,loqc}. If the beam splitter is symmetric the 
output state is
\begin{equation}
\hat{\rho}_{12}=\frac{1}{2}[|2\rangle_1 \langle2| |0\rangle_2\langle0|
+|0\rangle_1\langle0||2\rangle_2\langle2|]
\end{equation}
If the detector in arm 2 is perfect and no counts are recorded we know that two photons left in arm 1. 
Thus we can regard such a device as a two-photon state generator.

\begin{figure}[phtb]
\centerline {\epsfxsize=7.0cm \epsffile{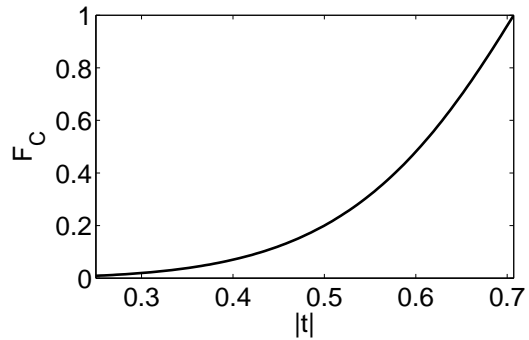}}
\caption {Correct output fidelity for a two-photon generator with loss in the beam splitter vs. transmission 
coefficient of the beam splitter ( $|t|=|r|$).}\label{lossypg}
\end{figure}

However if we have a lossy beam splitter which satisfies $|t|^2 + |r|^2 <1$ the output will contain other states such as 
$|1\rangle_1|0\rangle_2$ and $|0\rangle_1|0\rangle_2$ that also have a vacuum component in the 
detector arm, but do not produce the correct state in the other arm. The probabilities of 
producing the relevant states from this beam splitter are, from \cite{Barnett98,Jeffers00}, 
\begin{eqnarray}
p_{20}&=&  p_{02} =2|t|^2|r|^2  \nonumber\\
p_{10}&=& p_{01} =(|t|^2+|r|^2)(1-|t|^2-|r|^2)-(tr^*+rt^*)^2\nonumber\\
p_{00} & = & (1-|t|^2-|r|^2)^2+(tr^*+rt^*)^2
\end{eqnarray}
where $p_{nm}$ is the \textit{a priori} probability that the beam splitter produces the state 
$|n\rangle_1|m\rangle_2\langle m|_2\langle n|_1$, which depends on both the magnitude and phase of 
$t$ and $r$. When we measure 
zero photocounts in one arm the state in the other arm is,
\be
\hat{\rho}_1=\frac{p_{20}|2\rangle\langle2|+p_{10}|1\rangle\langle1|+p_{00}|0\rangle\langle0|}
{p_{20}+p_{10}+p_{00}}=P\ru{max}\hat{\rho}_c+\hat{\gamma}
\ee
Given a perfect detector, such that $P^R(n|n)=1\;\forall n$ we can equate $P\ru{max}$ with the correct 
output fidelity,
\begin{equation}
F\rs{c}=P\ru{max}=\frac{p_{20}}{p_{20}+p_{10}+p_{00}} \label{lbs}
\end{equation}
Even though the state produced by the lossless beam splitter is pure, this makes no difference to the form 
of the correct output fidelity. Eq. (\ref{lbs}) is plotted in Fig. (\ref{lossypg}) where we have assumed 
that $|t|=|r|$. This shows that the fidelity tends to unity as the beam splitter approaches 50/50. There 
is some freedom in the phase difference between reflection and transmission coefficients for a lossy beam 
splitter, ranging from zero for a perfect 50/50 to a $2\pi$ range for a $25/25$ or less, although, as can 
be seen by comparing the expressions for $p_{10}$ and $p_{00}$, the phase dependent terms in the 
denominator in Eq. (\ref{lbs}) cancel.
The general results in the previous section relating to imperfect detection also apply here. 
There is a reduction in fidelity because of non-unit quantum efficiency, which can be offset by 
preamplification, as has already been shown in this system \cite{Jeff07}.

\subsection{Nonlinear sign-shift gate}

\begin{figure}[hbp]
\centerline
{\epsfxsize=7.0cm \epsffile{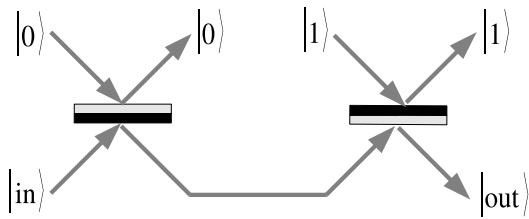}}
\caption{Nonlinear sign-shift gate \cite{Ralph01}. The beam splitter reflectivities are $|r_1|^2=5-3\sqrt{2}$ (left) and $|r_2|^2=(3-\sqrt{2})/7$ with phase changes of $\pi$ on reflection from the light grey sides.}
\label{NLSS}
\end{figure}

For the quantum optical gates which have been proposed as processing elements in quantum computers, gate outputs which have the correct states in the 
detector arms, but do not produce the correct state in the output arms can cause gate errors or project the system out of the computational space. This is a serious problem when only the perfect gate outout state performs the computational task. Then overlap-based fidelities will overestimate the gate fidelity \cite{Jeffers06}. A simple example of this is the nonlinear sign shift gate (see section V in \cite{Ralph01}) shown in Fig. (\ref{NLSS}). This gate makes the state transformation
\bea
\label{nlsseq}
\alpha |0\rangle +\beta |1 \rangle + \gamma |2 \rangle \rightarrow \alpha |0\rangle +\beta |1 \rangle - \gamma |2 \rangle,
\eea
and succeeds with probability $|r_2|^2=\left(3-\sqrt{2} \right)/7 \approx .2265$, where $r_2$ is the amplitude reflection coefficient of the second beam splitter. Two of these gates can be combined in parallel to form a two-qubit control-NOT gate \cite{Ralph01}. 
The fidelity of this gate was analysed in \cite{Jeffers06} for a lossy detector but perfect beam splitters, and it was indicated that the lower, retrodictive fidelity provides a good test of accurate device operation. If the beam splitters which make up the gate itself are lossy then the retrodictive fidelity becomes the correct output fidelity due to the inclusion of the extra factor $P\ru{max}<1$. 

If we assume that the transmission and reflection coefficients for both beam splitters are lowered by the same factor $K$, so that $|t_{1,2}|^2 +|r_{1,2}|^2 = K<1$, and that there is no phase change associated with the loss, the factor $P\ru{max}$ will be proportional to the loss of the two-photon component of the transformed state(Eq. (\ref{nlsseq})), as this will show the largest decrease. This component depends on the factor $r_1^2 \left(r_2^2-2t_2^2 \right)$, which is -1 for the device formed from lossless beam splitters. Thus this factor is decreased by $K^2$, and so $P\ru{max}=K^4$ - a fairly drastic reduction in fidelity even for relatively low loss.
\section{Conclusions}
In this paper we have introduced a measure of fidelity appropriate for postselecting devices which 
produce mixed states, generalising earlier pure state work \cite{Jeffers06}. For situations in which only a particular output state is useful the measure is especially appropriate, as it will normally 
be the probability that the device produces this useful output state. For this reason we have called this 
measure the correct output fidelity $F\rs{c}$. This is in contrast to more the normally used fidelity, 
Eq. (\ref{F}), which corresponds to the passing of a measurement test if one of the states is pure. $F\rs{c}$ forms a lower bound on 
this quantity. 

The correct output fidelity factorises into two parts, one of which depends 
only on the postselector design and components. These directly affect the 
output state produced when the detector functions perfectly. The second factor 
is based on the correct functioning of the detector, and is the probability 
that the detector correctly indicates the detector arm state. These two factors 
are the quantities that should be maximised in any postselecting device, and 
they will normally be simply expressible in terms of experimental quantities. This renders the correct output fidelity more simple to calculate than the normally used mixed state fidelity.

The results are illustrated using practical examples. In each case the 
output of the postselector is only useful when the correct state is output by 
the device, and so the correct output fidelity is an appropriate measure. For 
the comparison of coherent states the correct mixture of coherent states in the 
output arm occurs when two identical coherent states are chosen as inputs. The signature of this is the production of the vacuum state in the measurement arm. 
Poor detection efficiency increases the probability of obtaining no counts at 
a photodetector placed in this arm, and thus increases the probability that the 
state is incorrectly identified as the vacuum. One simple way around this is to 
preamplify the state input into the photodetector, which, at the cost of 
decreasing both the probability of obtaining no counts and the accessible 
information, increases the confidence in this result when it is obtained. 

The 
second example is that of two photon state generation using a lossy beam 
splitter. For a perfect 50/50 beam splitter in this case the output state would 
be pure, but the correct output fidelity measure still applies. 
The beam splitter is the basic element in all LOQC schemes, and if it is lossy this naturally impacts on gate fidelity.  Our final example, in which the beam splitters which form a nonlinear sign-shift gate are lossy, shows that the impact of the loss on the fidelity of such systems is considerable.

Until reliable push-button state makers become available postselection will be 
a major tool in quantum physics, and especially in optical implementations of 
quantum information systems. The standard measures of fidelity, for all their 
mathematical symmetry, are somewhat maladroit for such asymmetric circumstances, and can overestimate the usefulness of the states produced by 
postselectors. A fidelity measure such as the one described here overcomes this 
problem, providing a safe lower bound based on the simple criterion of whether 
or not the postselecting device performs the task that is required of it. 

\section*{Acknowledgments}
The authors would like to thank Erika Andersson, Igor Jex and Steve Barnett for useful discussions.

\end{document}